\documentclass[12pt,preprint]{aastex}  

\usepackage{epsfig}
\usepackage{natbib}

\shorttitle{Potential Vorticity Evolution of a Protoplanetary Disk}

\shortauthors{Li et al.}

\begin{document}

\title{Potential Vorticity Evolution of a Protoplanetary Disk with 
An Embedded Protoplanet}

\author{Hui Li\altaffilmark{1}, Shengtai Li\altaffilmark{2}, 
Josef Koller\altaffilmark{3}, Burton B. Wendroff\altaffilmark{2},
Richard Liska\altaffilmark{2,4}, Chris M. Orban\altaffilmark{5},  
Edison P.T. Liang\altaffilmark{6} and
Douglas N.C. Lin\altaffilmark{7}}
\altaffiltext{1}{
Applied Physics Division, MS B227, Los Alamos National
Laboratory, NM 87545; hli@lanl.gov}
\altaffiltext{2}{
Theoretical Division, MS B284, Los Alamos National
Laboratory, NM 87545; sli@lanl.gov; bbw@lanl.gov}
\altaffiltext{3}{
ISR Division, MS D466, Los Alamos National
Laboratory, NM 87545; jkoller@lanl.gov}
\altaffiltext{4}{Faculty of Nuclear Sciences and Physical Engineering,
Czech Technical University, Czech Republic; liska@siduri.fjfi.cvut.cz}
\altaffiltext{5}{
Department of Astronomy,
University of Illinois at Urbana-Champaign,
Urbana, IL 61801; corban@uiuc.edu}
\altaffiltext{6}{
Department of Physics \& Astronomy, Rice
University, Houston, TX 77005; liang@spacsun.rice.edu}
\altaffiltext{7}{UCO/Lick Observatory, University of California,
Santa Cruz, CA 95064; lin@ucolick.org}

\begin{abstract}
  We present two-dimensional inviscid hydrodynamic simulations of a
  protoplanetary disk with an embedded planet, emphasizing the
  evolution of potential vorticity (the ratio of vorticity to density)
  and its dependence on numerical resolutions. By analyzing the
  structure of spiral shocks made by the planet, we show that
  progressive changes of the potential vorticity caused by spiral
  shocks ultimately lead to the excitation of a secondary
  instability. We also demonstrate that very high numerical resolution
  is required to both follow the potential vorticity changes and
  identify the location where the secondary instability is first
  excited. Low-resolution results are shown to give the wrong
  location. We establish the robustness of a secondary instability and
  its impact on the torque onto the planet.  After the saturation of
  the instability, the disk shows large-scale non-axisymmetry, causing
  the torque on the planet to oscillate with large amplitude. The
  impact of the oscillating torque on the protoplanet's migration
  remains to be investigated.
\end{abstract}



\section{INTRODUCTION}

Well before extrasolar planets were discovered, Goldreich \& Tremaine
(1979; 1980) and Lin \& Papaloizou (1986a,b) have speculated that
tidal interactions between disks and embedded protoplanets would lead
to planet migration. Ward (1997) suggested that two different types of
migration could occur. One, called type I migration, is when the
protoplanet mass is still small enough that the migration rate can be
evaluated using linear analysis and is shown to be quite fast in the
sense that migration timescale is shorter than the estimated buildup
timescale of a giant protoplanet core. This has presented a serious
problem for the formation of giant planets. The other, called type II
migration, is when the protoplanet is massive enough that it opens
gaps in the disk. The protoplanet is then locked into the disk's
viscous evolution timescale, which is typically long.

The discovery of extrasolar planets with a large number of
short-period planets very close to their parent stars seems to support
a migration scenario since it is more likely that giant planet
formation takes place at larger distances outside the snow line. This
has generated renewed interests in the study of tidally induced planet
migration (e.g., Kley 1999; Lubow, Seibert, \& Artymowicz 1999; Nelson
et al. 2000; D'Angelo, Kley, \& Henning 2003). Both types of migration
have been basically confirmed by non-linear numerical simulations
though most simulations to-date are performed with a prescribed
viscosity.

A recent study by Balmforth \& Korycansky (2001) focused on the
evolution of potential vorticity in the co-orbital region of a planet
and they noted the possibility of secondary instabilities in the
corotation resonance region, which lead to the formation of
vortices. They showed that nonlinear dynamics of the co-orbital region
have an important influence on the saturation of the corotation
torque.

In an earlier paper \citep{KollerLiLin03}, we also investigated the
nonlinear dynamics in the co-orbital region (within $\sim 10$ Roche
lobe radii of the planet) through global nonlinear 2D inviscid
simulations. We found that the potential vorticity of the disk flow
undergoes systematic changes, presumably caused by the spiral shocks
generated by the planet though we did not give any detailed analysis
of the shocks. We also found that the flow eventually becomes unstable
due to the inflexion points in the potential vorticity profile.
Vortices are formed as a consequence.

This paper is a close follow up to Koller et al. Our purpose is two
fold. First, we present quantitative analyses of the ways that shocks
change the potential vorticity of the flow.  Second, through extensive
numerical resolution studies, we confirm the physical mechanism for
exciting a secondary instability through inflexion points in the
potential vorticity profile. However, we now find that the instability
is excited at a different location from what Koller et
al. presented. The phenomenon observed by Koller et al. is apparently
due to limited numerical resolutions, hence artificial. Nevertheless
we are still able to demonstrate the existence of such a secondary
instability, and its impact on the planet's torque. We have also
performed extensive numerical tests via different resolutions to check
the validity of these new findings.

This paper is organized as follows. We first describe our
numerical approach and how we set up our simulations in \S
\ref{sec:num} and \S \ref{sec:init}. We then present the analyses on
shocks and how they change the potential vorticity in \S
\ref{sec:shock}. The excitation of a secondary instability is
described in \S \ref{sec:s_instab}. The influence of numerical
resolution is presented in \S \ref{sec:resol}. Discussions are given
in \S \ref{sec:diss}.

\section{NUMERICAL MODEL}
\label{sec:num}

We assume that the protoplanetary disk is thin and can be described by
the inviscid two-dimensional isothermal Euler equations in a
cylindrical \{$r, \phi$\} plane with vertically integrated quantities.
The differential equations are the same as given in Kley (1998).
Simulations are carried out mostly using a split dimension hydro code
(Li \& Li 2005) whose basic algorithm is based on the MUSCL-Hancock
scheme (MHS) (Toro 1999). The methodology of MHS is extended here for
the presence of source terms with two following main modifications:
the fluxes are computed using primitive instead of conserved
variables, where van Leer flux limiter is applied to the slopes of the
primitive variables, and an iterative approximate Riemann solver is
used.  Standard dimensional splitting as well as some source splitting
are used. Radial sweeps with the angular flux derivatives omitted are
alternated with angular sweeps with the radial flux derivatives
omitted. All source terms except the planet's gravitational force are
included in the radial sweep. The planet force is applied in a
separate sweep as the numerical gradient of the planet potential, so
that the numerical curl of this gradient is zero, minimizing the
potential vorticity contamination by the planet.

We use the local co-moving angular sweep as proposed in the FARGO
scheme of Masset (2000) and modified in Li et al. (2001).  The idea is
basically to use a semi-Lagrangian form for the transport terms in the
angular sweep. A constant velocity is subtracted from the angular
transport velocity.  This velocity is chosen to be close to the mean
angular velocity, in order to move the data an integral number of
angular cells in one time step. This is then corrected by an integral
shift of the data.  This enables us to use a much larger time step than
would otherwise be possible.

The MHS scheme requires two ghost cells in the radial sweep (the
angular sweep is periodic). Holding these ghost cells at the initial
steady state produced the smallest boundary reflection of all boundary
conditions tried.

\section{INITIAL SETUP}
\label{sec:init}

The two-dimensional disk is modeled between $0.4 \leq r \leq 2$. The
planet is assumed to be on a fixed circular orbit at $r=1$. A
co-rotating frame is used and the positions of the central star and
the planet are fixed at $(r,\phi)=(0,0)$ and $(r_p,\phi_p)=(1,\pi)$
[acceleration due to frame rotation is also included, see Kley
(1998)].  The mass ratio between the planet and the central star is
$\mu=M_{p}/M_{\ast}$. Its Hill (Roche) radius is $r_H = r_p
(\mu/3)^{1/3}$. The disk is assumed as isothermal with a constant
temperature throughout the simulation region. The isothermal sound
speed, scaled by the Keplerian rotation speed $v_\phi$ at $r=1$, is
$c_s/v_{\phi}=H/r$ where $H$ is the disk scale height. Time is
measured by the orbital period at $r=1$.
  
We choose an initial surface density profile with $\Sigma(r) \propto
r^{-3/2}$, so that the ratio of vorticity to surface density,
potential vorticity (PV $\equiv \zeta$), has a flat radial profile
$\zeta = (\nabla\times {\bf v})_z/\Sigma \approx
\textrm{const}.\approx 0.5$. (Note that $\zeta$ has a small deviation
from being precisely $0.5$ due to the finite pressure gradient
which slightly modifies $v_\phi$ from its Keplerian value.)  With this
initial condition, we avoid the generation of inflection points due to
the rearrangement of $\zeta$ distribution as it is carried by the
stream lines \cite{bk01}.

We typically run the disk without the planet for 10 orbits first so
that the disk is settled into a numerical equilibrium (very close to
the analytic one). Then the planet's gravitational potential is
gradually ``turned-on'' over a 10-orbit period, allowing the disk to
respond to the planet potential gradually.  Furthermore, the planet's
potential is softened by an approximate three-dimensional treatment.
The basic idea is that, at any location \{$r,\phi$\}, the matter with
surface density $\Sigma$ is distributed (divided into many cells)
vertically over many scale heights (say $\pm 6H$) according to a
Gaussian isothermal profile. Radial and azimuthal forces exerted by
the planet at a specific \{$r,\phi,z$\} are calculated in this 3D
fashion. To get the 2D radial and azimuthal forces for a specific
\{$r,\phi$\}, one then integrates all the cells along $z$. Comparing
this pseudo-3D treatment with the usual 2D treatment where a fixed
smoothing distance is used, we find that it reduces the force close to
the planet (within $\sim 2 r_H$) by about a factor of 2 but converges
to the usual 2D force beyond $\sim 5 r_H$. Although there is some
accumulation of gas near the planet, we do not allow any disk gas to
be accreted onto it.  Self-gravity of the disk is not included.  Runs
are made using several radial and azimuthal grids to study the
influence of resolution, which ranges from $(n_{r}\times n_{\phi})=
200\times 800$ to $1200\times 4800$.  Simulations typically last
several hundred orbits at $r=1$.

One key feature of our simulations is that they are performed at the
inviscid limit, i.e., we do not explicitly include a viscosity term,
though numerical viscosity is inevitable and is needed to handle
shocks.  In addition, our simulations tend to be of higher resolution.
With a radial $n_r = 800$ (some runs go up to $n_r=1200, n_\phi=4800$)
and $0.4 \leq r \leq 2$, the diameter of the Hill sphere of a planet
with $\mu=10^{-4}$ is resolved by $\sim 32$ cells in each direction.

\section{SHOCK STRUCTURE and PV CHANGES}
\label{sec:shock}

\begin{figure}
\begin{center}
\epsfig{file=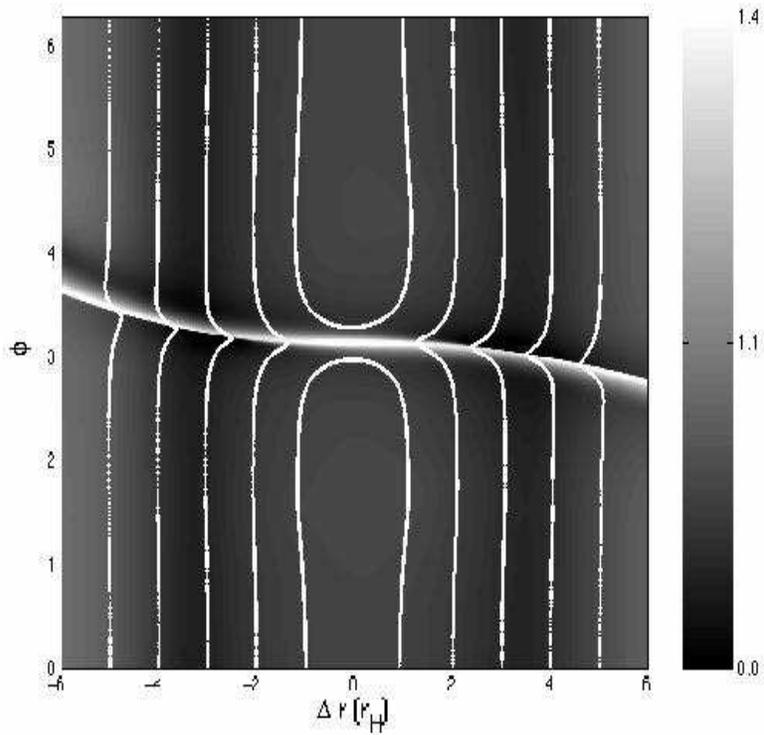,height=4in,width=4in,angle=0}
\end{center}
\caption{Surface density, multiplied by $r^{1.5}$, in the \{$r,\phi$\}
  plane at $t=100$ orbits for a run with $\mu=10^{-4}$, $c_s=0.05$,
  and $n_r\times n_{\phi} = 800\times 3200$. (The initial background
  surface density is unity, after multiplying $r^{1.5}$.) The planet
  is located at $\Delta r = 0$ and $\phi = \pi$. Two spiral shocks cut
  through the disk, bending stream lines. The PV changes induced by
  the shock are transported to different radii from their production
  site.}
\label{fig:den_strm}
\end{figure}

We will use one run in this section to facilitate discussion. This run
has $\mu=10^{-4}$, $c_s = 0.05$, $r_H = 0.0322$, and a $n_r\times
n_\phi = 800\times 3200$ resolution. Figure \ref{fig:den_strm} shows the
surface density (multiplied by $r^{1.5}$ so that the background
surface density is unity initially) in the \{$r,\phi$\} plane at
$t=100$ orbits. It shows the characteristic density enhancement at the
planet and at the two spiral shocks. Two density ``depressions'' at
$|\Delta r| \sim 3-4 r_H$ are developed as angular momentum is
gradually deposited there. Two density bumps (enhancements) occur
broadly over $|\Delta r| \sim 5-8 r_H$. Here, $\Delta r = r -
r_p$. Some streamlines (in the co-rotating frame) are plotted as well.

\subsection{PV Change by Shocks}

\begin{figure}
\begin{center}
\epsfig{file=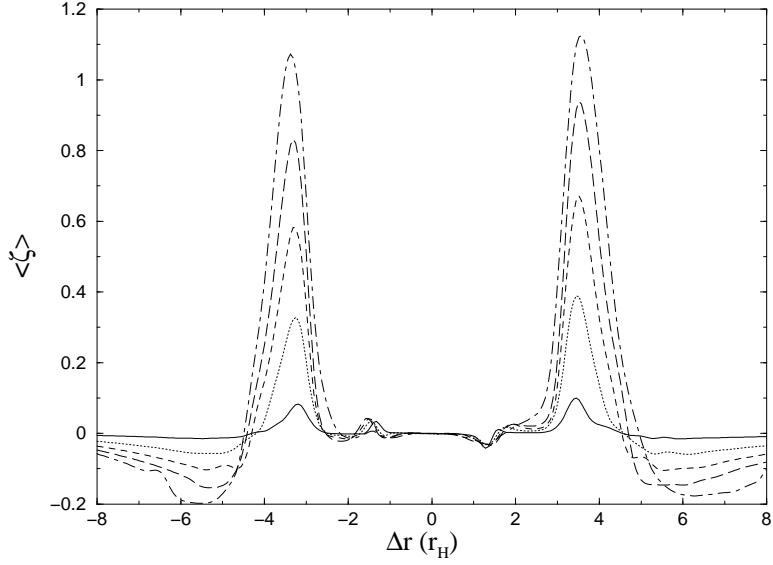,height=4in,width=3in,angle=-90}
\end{center}
\caption{Azimuthally averaged $\langle\zeta\rangle$ profile at
  $t=40~({\rm solid~ curve}), 100, 160, 220,$ and $300~({\rm dot-dashed~
  curve})$ orbits for a run with the same parameters as in Fig.
  \ref{fig:den_strm}. The planet is located at $\Delta r = 0$. The
  initial radial profile $\zeta \approx 0.5$ has been subtracted. Note
  the progressive growth of ``peaks'' and ``valleys'' in the
  $\langle\zeta\rangle$ profile. Also, $\langle\zeta\rangle$ in the
  planet region remains roughly unchanged.}
\label{fig:pv_r}
\end{figure}

Figure \ref{fig:pv_r} shows the azimuthally averaged
$\langle\zeta\rangle$ profile from $t= 40, 100, 160, 220,$ and $300$
orbits. (The initial value $\approx 0.5$ is subtracted.) Several
features can be seen. First, there are very small changes in
$\langle\zeta\rangle$ for $|\Delta r| \leq 2 r_H$. Second, at larger
distances from the planet, $\langle\zeta\rangle$ profile changes
progressively with time and shows both increase and decrease from its
initial value. An interesting feature is that even though both
``peaks'' and ``valleys'' are growing with time, two locations around
$\Delta r = \pm 4.5 r_H$ have no change in $\langle\zeta\rangle$.  A
natural question is what causes these features.

The flow around the planet can be separated into several sub-regions
depending on their streamline behavior (e.g., see Masset 2002),
including the horseshoe (or librating) region with $|\Delta r| \leq
r_H$, the separatrix region $r_H < |\Delta r| < \sqrt{12} r_H$, and
the streaming region $|\Delta r| > \sqrt{12} r_H$. Typically, spiral
shocks do not occur until the relative velocity between the disk flow
and the planet becomes larger than the sound speed. This roughly
translates to $|\Delta r| \approx H$. So, a critical parameter is
$H/r_H \approx c_s/r_H$, which gives an indication whether the spiral
shock is starting from within the planet's Roche lobe or not
(see also Korycanski \& Papaloizou 1996). In the limit
of small $c_s/r_H$, flow around the planet becomes quite complicated
(e.g., see Tanigawa \& Watanabe 2002). Another added yet potentially
very important complication is that any numerical error coming out
nearing the planet might be strongly amplified by these shocks, giving
numerical artifacts which can be physically unreal. So, we have mostly
used a relatively higher sound speed of $c_s=0.05~ (> r_H)$ in this
study. Higher sound speeds mean weaker shocks, which typically also
mean slower changes in the disk properties.

As discussed in Koller et al. (2003), spiral shocks are capable of
destroying the conservation of PV. For the parameters used in this
run, we do not expect shocks within $\sim \pm 2 r_H$, which is
confirmed by the fact that $\langle\zeta\rangle$ stays roughly at the
initial value. Once in the shocked region, we can use a very useful
formula for calculating the PV jump across a (curved) shock, which is
given in Kevlahan (1997), after some manipulation,
\begin{equation}
\label{eq:dzeta}
\Delta \zeta = \frac{(M_\perp^2-1)^2}{M_\perp^2} \frac{\partial
M_\perp}{\partial \tau}\frac{c_s}{\Sigma_2}~,
\end{equation}
where $M_\perp$ is the perpendicular Mach number ($u_\perp/c_s$) which
is equal to $(\Sigma_2/\Sigma_1)^{1/2}$ for isothermal
shocks. $\Sigma_1$ and $\Sigma_2$ are the pre- and post-shock surface
densities. The direction $\tau$ is tangential to the shock front and
is defined to be pointing away from the planet. According to
Eq. (\ref{eq:dzeta}), the sign of $\Delta \zeta$ is determined only by
the gradient of $M_\perp$ along the shock front.

\begin{figure}
\begin{center}
\epsfig{file=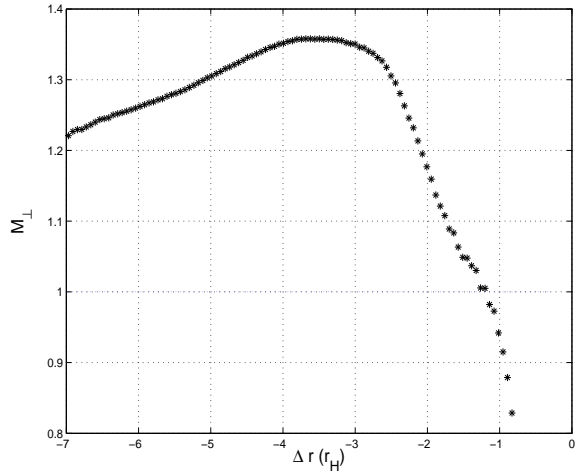,height=2.5in,width=3in,angle=0}
\end{center}
\caption{Perpendicular Mach number $M_\perp$ along the spiral shock front
  for the run shown in Fig. \ref{fig:den_strm} at $t=100$ orbits. Note
  that the radial distances here mark the PV production sites, to
  which systematic shifts shown in Fig. \ref{fig:den_strm} have to be
  added when compared with the radial locations in
  Fig. \ref{fig:pv_r}.}
\label{fig:machperp}
\end{figure}

Figure \ref{fig:machperp} shows $M_\perp$ as a function of radial
distance to the planet (only one side is shown). This is obtained by
first identifying the high compression region through calculating the
velocity compression. After finding the flow impact angle to that
surface, one can then calculate $M_\perp$ at the
shock. Differentiating along that surface gives $\partial
M_\perp/\partial \tau$. This figure shows the general behavior of what
one might expect of a spiral shock produced by the planet. As one
moves radially away from the planet, the flow goes from a non-shocked
state to having shocks, which means that $M_\perp$ gradually increases
from being $< 1$ to $1$ when the shock actually starts. Very
far away from the planet, the pitch angle between the spiral wave and
the background flow becomes small enough that $M_\perp$ is expected to
become less than one again. So, as a function of radial distance from
the planet, $M_\perp$ is expected to increase first, reaching its
peak, then starting to decrease, it eventually drops to be smaller than
one at some large distance. Consequently, using eq. (\ref{eq:dzeta}),
we can expect that $\Delta \zeta$ should be positive first, becomes
zero when $M_\perp$ reaching its peak, then is negative when $M_\perp$
is decreasing. This matches quite well with the behavior of
$\langle\zeta\rangle$ given in Fig. \ref{fig:pv_r}. Note that over the
whole evolution, shocks are quite steady with very slow evolution,
indicated by the small ``outward'' movement of the inflexion points
(where $\Delta \zeta
\approx 0$). All these features are quite generic and we have
confirmed this by using several different planet masses. 

From Fig. \ref{fig:machperp}, we can see that $M_\perp$ peaks at
$\Delta r \approx -3.8 r_H$, but the point where $\Delta \zeta = 0$ is
at $-4.5 r_H$ based on Fig. \ref{fig:pv_r}.  To quantitatively match
the two profiles in terms of radial distances to the planet, one has
to take into the account the fact that the radial location where shock
occurs (and the flow's PV is changed) is {\em different} from where
these PV changes are eventually deposited. This is illustrated in Fig.
\ref{fig:den_strm}.  Correcting for this systematic shift in radial
locations, the agreement between the azimuthally averaged
$\langle\zeta\rangle$ profile and the $M_\perp$ profile is quite good.

To summarize, the spiral shocks emanating close to the planet cause
systematic changes to the PV distribution of the disk flow. As the
disk rotates, the disk flow repeatedly passes through these shocks so
that changes in PV then accumulate. The sign of PV change is
determined by the shock structure, and the amplitude of change is
determined both by the shock structure [eq. (\ref{eq:dzeta})] and how
many times the flow passes through the shocks in a given time [$\sim
(r^{-3/2} -1) \Delta t/2\pi$]. We have identified the physical cause
for having both positive and negative changes in PV, in addition to
the existence of an inflexion point. We defer the quantitative
comparison of the amplitude of PV changes between simulations and eq.
(\ref{eq:dzeta}) to a future study.

\section{SECONDARY INSTABILITY}
\label{sec:s_instab}

\subsection{Instability and Vortices}

Having established the profile of $\langle\zeta(r,t)\rangle$, we now
turn to the issue of secondary instability. This type of instability
is usually associated with the existence of inflexion points (extrema)
in the PV profile (see Drazin \& Reid 1981) and often requires
satisfying a threshold condition. As shown in Fig. \ref{fig:pv_r}, the
PV profile has extrema points, and the magnitude of the slope
between the peak at $\Delta r \approx 3.5 r_H$ and the valley at
$\Delta r \approx 4.6 r_H$ is increasing with time (similarly on the
other side of the planet). It is then expected that these changes will
become large enough that a threshold is reached and secondary
instabilities will be excited.

To demonstrate this, we have made runs out to several hundred orbits
($400-800$ orbits depending on resolution). Figure \ref{fig:pv_2d} shows
the evolution of PV in the \{$r,\phi$\} plane for a run with $\mu =
10^{-4}$, $c_s = 0.05$, and $n_r\times n_\phi = 400\times
1600$. Indeed, a secondary instability is observed to occur in the
outer ``valley'' of the PV profile at $t \approx 180$ orbits. (The
other valley in the $\sim -5 r_H$ region also becomes unstable at a
later time.)  The negative PV regions correspond to the two density
bumps, which are roughly axisymmetric early on. When the instability
is excited, density bumps break up into $\sim 5$ anticyclonic vortices
(even higher density blobs). Over the next tens of orbits, as the
instability evolves and saturates, these density blobs merge, forming
density enhancements that are non-axisymmetric on large scales. This
is illustrated in Fig. \ref{fig:den} at $t=500$ orbits. Note that
density depression regions also become non-axisymmetric.

\subsection{Torque Evolution}

As discussed in Koller et al. (2003), one primary goal of studying
secondary instability is to understand its impact on the torque
evolution of disk gas on the planet. Figure \ref{fig:torque} shows the
torque history on the planet for a run with the same parameters as in
Figure \ref{fig:pv_2d}. The rapid oscillations in the torque history
starting at $t \approx 180$ indicates the initiation of the
instability (consistent with Fig. \ref{fig:pv_2d}). As the higher
density blobs pass by the planet, they exert stronger torques.
Furthermore, Figure \ref{fig:torque} shows that the oscillation
amplitude grows with time, which is caused by the gradual density
increase in the PV ``valleys'' over long timescales. To better
understand the origin of fast oscillations, we have calculated the
torques separately from three different parts of the disk, which are
inner: $\Delta r < 3 r_H$, middle: $|\Delta r| \leq 3 r_H$, and outer:
$\Delta r > 3 r_H$, respectively. Figure \ref{fig:torq_det} shows
their individual contributions over a time interval from $t=490 -
512$. We can see that both inner and outer regions have a periodic
large amplitude variation. These are caused by the orbital motion of
the non-axisymmetric disk flow at $\Delta r \sim -5 r_H$ and $\Delta r
\sim 7.5 r_H$, respectively (see Fig. \ref{fig:den}). In fact, the
torque oscillation periods from these two regions are precisely the
expected orbital periods at those radii. Since these two periods are
different, their sum gives an erratic appearance (Fig.
\ref{fig:torque}).

One might speculate that even with these large amplitudes and rapid
oscillations, there seems to be a mean value that is negative and has
a small amplitude, consistent with the type I migration
expectations. But since our simulations have the planet moving on a
fixed circular orbit, it is premature to conclude that these
oscillations will not change the type I migration picture. Studies
allowing the planet to migrate under the influence of these
oscillating torques is under way to address this important issue.

Here, we have mostly concentrated on demonstrating the existence of
such a secondary instability and its potential impact on the planet's
torque. We have not quantitatively analyzed what is the threshold
condition in the PV profile that excites this instability. Such a
threshold obviously depends on the planet mass, shock structure, sound
speed of disk gas, and viscosity in the disk, etc. We have assumed the
inviscid limit. Sufficient viscosity could potentially remove the
buildup of the PV's peaks and valleys made by shocks, hence never allowing
the profile to reach the critical destabilization level.

\section{RESOLUTION STUDY}
\label{sec:resol}

The results we presented here, especially Figs. \ref{fig:pv_r} and
\ref{fig:pv_2d}, are different from the findings in Koller et al.
(2003, see their Fig. 3, also curve A in our Fig.
\ref{fig:pv_resolu}). Results there showed additional ``dips''
interior (i.e., closer to the planet) to the main positive peaks in
$\langle\zeta\rangle$. Furthermore, those ``dips'' were shown to
deepen with time and eventually became unstable.  

Figure \ref{fig:pv_resolu} shows a comparison of runs with 4 different
resolutions but the same set of physical parameters. Even though
different curves of $\langle\zeta\rangle$ converge at large $\Delta r$
($> 6 r_H$) because the shock is weak and the gradient is small there,
the behavior at small $\Delta r$ (where the shock is strong) shows
large differences. Comparing curves A and C, for example, we can see
that when the resolution is low and shocks are not well resolved, a
larger part of the disk was affected (see the region of $\Delta r \sim
0.5 - 2.5 r_H$ in curve A). We were able to reproduce the results of
Koller et al. in our low-resolution runs but our high-resolution
studies indicate that the ``dips'' observed in Koller et al. are
getting narrower and shallower with higher resolutions.  This means
that the previous results showing the inner ``dips'' are numerical
artifacts, not true physical effects. Note that the PV within the
planet's Roche lobe is not conserved due to both the switch-on of the
planet's mass (no matter how slowly) and the numerical error in
implementing the planet's potential. This is indicated by Fig.
\ref{fig:pv_two}, where PV in the \{$r,\phi$\} plane at $t=100$ is
shown for two different resolutions. Small PV changes are visible
coming out of the planet region. For the low resolution case, the PV
change from the planet is ``spread'' over a wide region. It is then
amplified by poorly resolved shocks.  Such amplification eventually
leads to an instability, unfortunately, in much the same spirit we
have discussed here. For the high resolution run, the shock and the PV
change from the planet is well separated. The erroneous amplification
does not occur.

It is seen that very high resolutions ($800\times 3200$ to $1200\times
4800$) are needed to obtain convergence in the $\langle\zeta\rangle$
profile (curves C and D).  We emphasize that even though there are
still some minor differences for the high resolution runs we presented
here, the overall feature of having a positive peak and a negative
valley is quite consistent among all resolutions. Furthermore, the
instability now comes out of the valley region, which typically has a
much better convergence than other parts of the $\zeta$ profile.  In
addition, we have also made a run using pure Lax-Wendroff scheme with
$n_r \times n_\phi = 800\times 3200$ (curve E). Comparing with curve
C, it gives very similar results. It also shows the instability at
later times (not shown here), though the MHS scheme gives sharper and
smoother shocks than hybrid schemes.


\section{CONCLUSIONS}
\label{sec:diss}

We have carried out high-resolution two-dimensional hydrodynamic disk
simulations with one embedded protoplanet.  We find that the total
torque on the planet, caused by tidal interactions between the disk
and the planet, can be divided into two stages. First, it is negative,
and slowly varying, consistent with the type I migration expectation.
Second, it shows large amplitude and very fast oscillations, due to
the excitation of an instability which first breaks up the
axisymmetric density enhancement into higher density blobs, e.g.,
vortices.  These vortices then merge, forming large-scale
non-axisymmetric density enhancements. This non-axisymmetry causes the
torque to continuously oscillate.

In Koller et al. (2003), we were misled by a spurious feature in the PV
profile, due to an inadequate numerical resolution, which eventually
became unstable. Despite the inaccurate location where a secondary
instability is initiated, the physical explanation proposed by Koller
et al. for exciting such an instability is well founded.  We now have
a self-consistent picture from performing very high resolution
simulations to understand the shock structure and the PV profile.
Further studies allowing the planet to migrate under the influence of
these torques will be necessary and very interesting.

\acknowledgments
This research was performed under the auspices of the Department of
Energy. It was supported by the Laboratory Directed Research and
Development Program at Los Alamos and by LANL/IGPP.

\begin{figure}
\begin{center}
\end{center}
\caption{PV in the \{$r,\phi$\} plane at $t=160, 180, 200,$ and $240$
  orbits (panels A-D) with $n_r\times n_{\phi} = 400\times 1600$.
  (Results are similar for other resolutions.) Panels B and C show the
  initiation of a secondary instability in the ``valley'' regions with
  $|\Delta r| = 5-7 r_H$.  This instability breaks the flow into
  higher density blobs (``vortices'') which are indicated by the
  isolated PV depressions (panels B and C).  These vortices merge at
  late times, forming large-scale asymmetries in the azimuthal
  direction (panel D) .}
\label{fig:pv_2d}
\end{figure}
\begin{figure}
\begin{center}
\end{center}
\caption{Surface density, multiplied by $r^{1.5}$, in the \{$r,\phi$\}
  plane at $t=500$ orbits. (Initial surface density is unity at $r=1$,
  after multiplying $r^{1.5}$.) After the instability has saturated,
  the density distribution becomes strongly non-axisymmetric. When
  they move around the disk, the torque on the planet oscillates.}
\label{fig:den}
\end{figure}
\begin{figure}
\begin{center}
\epsfig{file=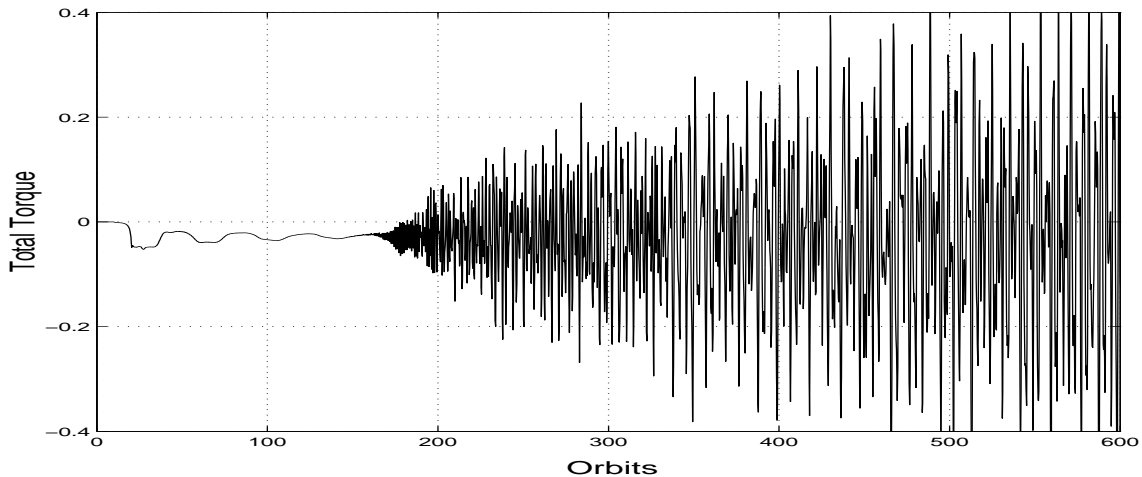,height=2.5in,width=6in,angle=0}
\end{center}
\caption{The total torque on the planet vs. time from a run having the same
  parameters in Fig. \ref{fig:pv_2d}. Note that once the flow becomes
  unstable after $t=180$, the total torque shows large amplitude
  oscillations.}
\label{fig:torque}
\end{figure}
\begin{figure}
\begin{center}
\epsfig{file=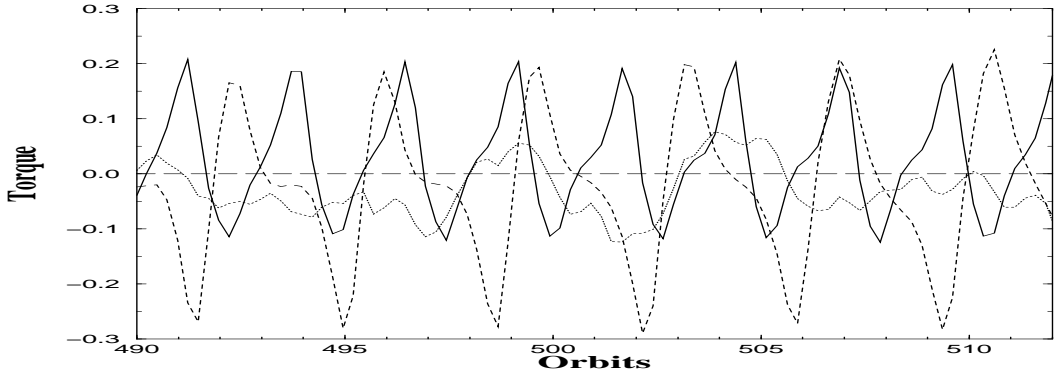,height=5.5in,width=2.in,angle=-90}
\end{center}
\caption{A zoomed-in and ``decomposed'' view of Fig. \ref{fig:torque} 
  during orbits $t=490-512$. The solid, dotted and dashed curves are
  torques from three different parts of the disk: $\Delta r < 3 r_H$,
  $|\Delta r| \leq 3 r_H$, and $\Delta r > 3 r_H$, respectively. The
  periodic oscillations in torque profiles (solid and dashed curves)
  are caused by a disk with non-axisymmetric density profiles rotating
  around the planet (see Fig. \ref{fig:den}).  When the higher density
  region moves close to the planet, it gives a stronger interaction.
  The oscillation periods in torques are the same as their respective
  orbital periods.}
\label{fig:torq_det}
\end{figure}
\begin{figure}
\begin{center}
\epsfig{file=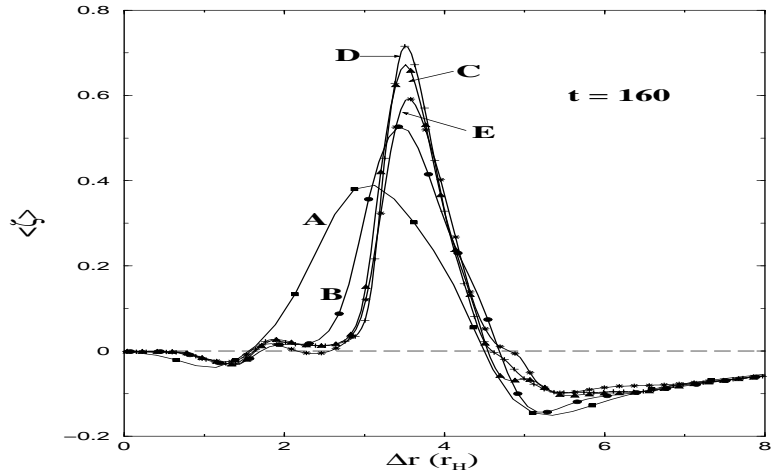,height=2.5in,width=4in,angle=0}
\end{center}
\caption{Azimuthally averaged $\langle\zeta\rangle$ profile at $t=
160$ orbits with $\mu=10^{-4}$ and $c_s=0.05$, but different
resolutions: curves (A-D) use the MHS scheme with $n_r\times n_{\phi}
= 200\times 800, 400\times 1600, 800\times 3200, 1200\times 4800$,
respectively. Curve E uses the pure Lax-Wendroff scheme with
$n_r\times n_{\phi} = 800\times 3200$. Curves C and D are quite
similar to each other, indicating convergence. }
\label{fig:pv_resolu}
\end{figure}
\begin{figure}
\begin{center}
\end{center}
\caption{PV in the \{$r,\phi$\} plane at $t=100$ orbits with
$n_r\times n_{\phi} = 200\times 800$ (left) and $800\times 3200$
(right). The planet is located at $\Delta r = 0$ and $\phi =
\pi$. Only one side is shown. Resolving the shock is important in
determining the starting point of the shock and where the PV changes
will be deposited.}
\label{fig:pv_two}
\end{figure}

\begin{thebibliography}{14}

\bibitem[Balmforth \& Korycansky 2001]{bk01}
{Balmforth}, N.~J. \& {Korycansky}, D.~G. 2001, \mnras, 326, 833


\bibitem[D'Angelo, Kley, \& Henning 2003]{DAngeloKleyHenning03} D'Angelo,
G., Kley, W., \& Henning, T.\ 2003, \apj, 586, 540

\bibitem[{{Drazin} \& {Reid}(1981)}]{DrazinReid81}
{Drazin}, P. G. \& {Reid}, W.H. 1981, {\it Hydrodynamics Stability},
Cambridge

                                          
\bibitem[Goldreich \& Tremaine (1979)]{GoldreichTremaine79}
{Goldreich}, P. \& {Tremaine}, S. 1979, \apj, 233, 857          

\bibitem[Goldreich \& Tremaine (1980)]{GoldreichTremaine80}
--- 1980, \apj, 241, 425              


\bibitem[Kevlahan 1997]{kev97}
Kevlahan, N. K.-R. 1997, J. Fluid Mech., 341, 371

\bibitem[Kley 1998]{Kley98}
{Kley}, W. 1998, \aap, 338, L37

\bibitem[Kley 1999]{Kley99}
{Kley}, W. 1999, MNRAS, 303, 696

\bibitem[Koller et al. 2003]{KollerLiLin03}
Koller, J., Li, H., \& Lin, D.N.C. 2003, \apj, 596, L91-94 

\bibitem[{{Korycanski} \& {Papaloizou} (1996)}]{KorycanskyPapaloizou96}
{Korycansky}, D. \& {Papaloizou}, J.C.B. 1996, \apjs, 105, 181

\bibitem[Li et al. 2001]{lietal01}
{Li}, H., {Colgate}, S.~A., {Wendroff}, B., \& {Liska}, R. 2001, \apj,
551, 874

\bibitem[Li \& Li 2005]{lili05} Li, S., \& Li, H. 2005, \apjs, to be
submitted

\bibitem[Lin \& Papaloizou 1986a]{LinPap86a}
Lin, D.~N.~C. \& Papaloizou, J. 1986a, \apj, 307, 395

\bibitem[Lin \& Papaloizou 1986{\natexlab{b}}]{LinPap86b}
--- 1986b, \apj, 309, 846

\bibitem[Lubow, Seibert, \& Artymowicz 1999]{lub99}
Lubow, S.H., Seibert, M. Artymowicz, P. 1999, \apj, 526, 1001

\bibitem[Masset 2000]{Masset00}
{Masset}, F.~S. 2000, \aaps, 141, 165


\bibitem[Masset (2002)]{Masset02}
--- 2002, \aap, 387, 605

\bibitem[Nelson et~al.2000]{NelsonPapMassetKley00}
{Nelson}, R.P., {Papaloizou}, J.C.B., {Masset}, F., \& {Kley}, W. 2000,
  \mnras, 318, 18



\bibitem[Tanigawa \& Watanabe 2002]{tw02}
Tanigawa, T., \& Watanabe, S. 2002, \apj, 580, 506
                                
\bibitem[Toro 1999]{toro99}
Toro, E.F. 1999, Riemann Solvers and Numerical Methods for
  Fluids Dynamics, Springer, Berlin, Heidelberg, Second Edition

                                
\bibitem[{{Ward}(1997)}]{Ward97}
{Ward}, W.~R. 1997, Icarus, 126, 261

\end{thebibliography}

\end{document}